\documentclass[12pt]{article}
\usepackage[dvips]{graphics}
\usepackage{amssymb}
\newcommand{\be}{ \begin{equation} }
\newcommand{\ee}{\end{equation}} 
\begin{document} 
\def\theequation{\arabic{section}.\arabic{equation}} 
\begin{titlepage} 
\title{Phase space geometry in scalar-tensor cosmology} 
\author{Valerio Faraoni\\ \\ 
{\small \it Physics Department, Bishop's University}\\ 
{\small \it Lennoxville, Qu\`{e}bec, Canada J1M~1Z7}\\\\
{\small and}\\\\
{\small \it Physics Department, University of Northern 
British Columbia}\\ 
{\small \it 3333 University Way, Prince George, B.C., Canada V2N~4Z9}\\ 
{\small \it  email~~vfaraoni@unbc.ca} }
\date{} \maketitle 
\vspace*{1truecm} 
\begin{abstract} 
We study the phase space of spatially homogeneous and isotropic cosmology in general
scalar-tensor theories. A reduction to a two-dimensional phase space is performed 
when possible --- in these situations the phase space is usually a two-dimensional 
curved surface embedded in a 
three-dimensional space and composed of two sheets attached to each other, possibly 
with complicated topology. 
The results obtained are independent of the choice of the coupling function 
of the theory and, in certain situations, 
also of the potential.
\end{abstract} 
\vspace*{1truecm} 
\end{titlepage}

\def\theequation{\arabic{section}.\arabic{equation}}


\section{Introduction}
\setcounter{equation}{0}

Brans-Dicke and scalar-tensor gravity have 
been studied for over four decades. The original motivation for the 
introduction of Brans-Dicke theory \cite{BD} came from Mach's principle 
and cosmology but since Brans and Dicke's original paper, significant 
motivation has been added for research in scalar-tensor gravity. First, 
Brans-Dicke  theory has been generalized to scalar-tensor 
theories with coupling functions instead of coupling constants  \cite{ST}.
More important, it has been realized that scalar-tensor gravity contains 
features that are common to supergravity and string or M-theory.
The low-energy limit of the bosonic string theory reduces to a Brans-Dicke  
theory 
with Brans-Dicke parameter $\omega=-1$ \cite{string}.

Renewed interest in cosmology in scalar-tensor gravity (see Refs. 
\cite{mybook,FujiiMaeda})
has been motivated by the extended \cite{extended} and hyperextended 
\cite{hyperextended} scenarios of inflation in the early universe and by 
scalar-tensor 
models of dark energy trying to explain the present acceleration of the universe 
\cite{STquintessence}.
Even with the simplifying symmetry of spatially homogeneous and isotropic 
Friedmann-Lemaitre-Robertson-Walker 
(hereafter ``FLRW'') cosmology, the field equations of 
scalar-tensor gravity for the metric and the Brans-Dicke-like scalar field are 
complicated, only a few exact solutions are available, and a phase space picture 
is very useful for the qualitative description of the dynamics. A few 
excellent 
and detailed works 
on the dynamics of Brans-Dicke homogeneous and isotropic cosmological 
models are available 
\cite{KE}--\cite{HW}. The possibilities studied in the literature 
include giving a mass or other potential $V( \phi)$ to the 
Brans-Dicke-like field 
$\phi$ 
(as is natural in particle physics), adding a perfect fluid while keeping 
$V \equiv 0$, adding a 
cosmological constant $\Lambda$, and/or considering 
FLRW spaces with the three
possible values $0, \pm 1$ of the curvature index $K$. However, the geometric 
structure of the phase space, including its dimensionality, is still poorly 
understood for scalar-tensor theories more general than Brans-Dicke 
theory. 
In addition, the studies of the dynamics available in the literature often
use variables that mix the scale factor $a(t)$ of the FLRW metric with the
Brans-Dicke-like field $\phi$. Although convenient from the formal point 
of view, the 
use of these variables tends to obscure the physical picture and it 
may be difficult to
rephrase results in terms of the physical variables traditionally used in cosmology,
such as the scale factor $a$, the Hubble parameter $H$, the comoving time $t$, and 
the scalar field $\phi$.

In this paper we focus on spatially homogeneous and isotropic FLRW 
spaces described by the metric
\be  \label{1}
ds^2=-dt^2+a^2(t)\left[ \frac{dr^2}{1-Kr^2} +r^2 \left( d\theta^2 +\sin^2 \theta \, 
d\varphi^2 \right) \right]
\ee   
in comoving coordinates $\left( t,r, \theta, \varphi \right)$ and with curvature 
index $K$. We consider scalar-tensor 
theories of gravity described by the action
\be  \label{2}
S^{(ST)}=\frac{1}{16\pi} \int d^4 x \, 
\sqrt{-g} \left[ \phi R -\frac{\omega(\phi)}{\phi} \, g^{ab} \nabla_a \phi\, \nabla_b \phi 
-V(\phi) \right] +S^{(m)} \;,
\ee
where $S^{(m)}$ is the action describing ordinary matter, $g$ is the determinant of the metric tensor $g_{ab}$,
$\nabla_c$ is the covariant derivative operator associated with $g_{ab}$, $R$ is 
the Ricci curvature of spacetime, and 
$\phi$ is the gravitational Brans-Dicke-like scalar self-interacting 
through the potential $V(\phi)$.
The metric signature is $-,+,+,+$, 
$\Box \equiv g^{bc} \nabla_b\nabla_c $, and we follow the notations and 
conventions of Ref.~\cite{Wald}.

The study of FLRW cosmology 
at very early times has limited value because isotropization
is probably achieved by inflation, which washes away information about the pre-inflationary state of the universe. At 
very early times before inflation the universe may have been anisotropic 
(see Refs. \cite{Jorge} for studies of isotropization of Bianchi 
models in scalar-tensor gravity). However,
it is meaningful 
to consider FLRW universes during and 
after inflation and in particular during the present
and future evolution of the universe. In the quest for successful  models
of dark energy able to explain the present acceleration of the universe 
deduced from the study 
of type Ia supernovae \cite{SN}, scalar-tensor gravity has been used many times
 \cite{STquintessence}. In this context, it is suggested \cite{Starobinsky} that
the universe may evolve into a chaotic regime, thus making it 
impossible to predict its evolution in the distant future. 
The dimensionality of the phase space is important in this regard because a dimensional reduction of the phase 
space to two dimensions (which is achieved in 
this paper for many situations in the context of
general scalar-tensor theories) is believed to 
inhibit the presence of chaos (see Ref.~\cite{Gunzigetal}
 for the special case of a nonminimally coupled scalar field theory).
The impossibility of chaos in a two-dimensional phase 
space is guaranteed when the phase space is flat but it is not entirely trivial 
when the phase space is a curved two-dimensional surface, 
as is often the case in scalar-tensor cosmology \cite{inprogress}.

In this paper we do not provide a complete phase space analysis as is done
in previous works for Brans-Dicke theory and for special choices of the 
scalar field potential $V(\phi)$ and of the coupling function $\omega(\phi)$ 
\cite{KE}--\cite{HW}, but we focus instead on the geometric structure 
and dimensionality of the phase space for any coupling function  
$\omega(\phi)$ and (except special cases) for any potential $V(\phi)$. A complete
phase space analysis  cannot obviously  be performed without 
specifying the functions $\omega$ and $V$, but we believe that 
a unified treatment of the phase space geometry in the general case is of 
interest, especially in view of the discussion of chaotic dynamics and of 
the uncertainty in the form of $\omega$ and $V$. 
Results similar to those obtained in this paper have been presented  elsewhere 
for a special scalar-tensor theory \cite{Foster}--\cite{SaaetalIJTP}, that 
of a scalar field
coupled nonminimally with the Ricci curvature and described by the action
\be  \label{NMC}
S^{(NMC)}=\int d^4x \, \sqrt{-g} \left[ \left( \frac{1}{2\kappa}
-\frac{\xi}{2} \phi^2 \right) R -\frac{1}{2}
 g^{ab}\nabla_a\phi \, \nabla_b \phi -V(\phi) \right] +S^{(m)} \;,
\ee
where $\kappa=8\pi G$ and $\xi$ is a dimensionless coupling 
constant (in our notations $ \xi=1/6$ corresponds to conformal coupling).
This theory is formally a scalar-tensor theory. In fact,
by redefining the scalar field according to 
\be
\varphi \equiv \frac{1-\kappa\xi \phi^2}{G} \;,
\ee
the action (\ref{NMC}) is written as 
\be
S^{(NMC)}=\int d^4x \, \frac{\sqrt{-g}}{16\pi} \left[ 
\varphi \, R -\frac{\omega(\varphi)}{\varphi}
 g^{ab}\nabla_a\varphi \, \nabla_b \varphi -U(\phi) \right] +S^{(m)} \;,
\ee
where
\be
\omega(\varphi) =\frac{G\varphi}{4\xi \left(1-G\varphi \right)} 
\ee
and
\be
U(\varphi )= 16\pi V\left[ \phi ( \varphi) \right] =
16\pi V\left( \pm \sqrt{ \frac{1-G\varphi}{8\pi G\xi} } \right) \;.
\ee
The theories in which the geometry and the dimensionality 
of the phase space of scalar-tensor FLRW cosmology are known are 
summarized in Table~1. 
This table summarizes also the main results of the present paper, 
which are derived in the following sections.

\section{The equations of scalar-tensor cosmology }
\setcounter{equation}{0}

In this section we consider scalar-tensor gravity as described by the action
(\ref{2}) and we write the field equations in a form that is convenient for 
later use. The field equations obtained by varying the action (\ref{2}) 
are
\be \label{3}
G_{ab}=\frac{\omega( \phi)}{\phi^2} \left[ \nabla_a \phi \, \nabla_b \phi 
-\frac{1}{2} \, g_{ab} \,\nabla^c \phi \, \nabla_c \phi
\right] + \frac{1}{\phi} \left( \nabla_a \nabla_b \phi -g_{ab} 
\Box \phi \right) 
-\frac{V}{2\phi} \, g_{ab} +\frac{8\pi}{\phi} \, T_{ab}^{(m)} \;,
\ee

\be  \label{4}
\Box \phi=\frac{1}{2\omega+3}\left( \phi \, 
\frac{dV}{d\phi} -2V -\,\frac{d\omega}{d\phi}
\nabla^c\phi \, \nabla_c\phi +8\pi T^{(m)} \right) \;,
\ee
where $T_{ab}^{(m)}$ is the stress-energy tensor of ordinary 
matter (usually treated as a fluid) which is covariantly conserved, 
$\nabla^b T_{ab}^{(m)}=0$, and $T= {T^a}_a$. Note that the combination
$\phi dV/d\phi -2V$ in eq.~(\ref{4}) disappears if the potential consists of a
 pure mass term $V(\phi)=m^2\phi^2/2$ (which is the special case 
considered in Ref.~\cite{SG} for Brans-Dicke theory). In the FLRW metric 
(\ref{1}) the field 
equations assume the form
\be  \label{5}
H^2 =-H\left( \frac{\dot{\phi}}{\phi}\right) +\frac{ \omega(\phi)}{6} 
\left( \frac{\dot{\phi}}{\phi} \right)^2 +\frac{ V(\phi)}{6\phi} -\frac{K}{a^2} 
+\frac{8\pi \rho^{(m)}}{3\phi} \;,
\ee
\begin{eqnarray}  
\dot{H} &= & -\frac{ \omega(\phi)}{2} \left( \frac{\dot{\phi}}{\phi}\right)^2 
+2H \left( \frac{\dot{\phi}}{\phi} \right) +\frac{1}{2\left( 2\omega+3 \right)\phi}
\left[ \phi\frac{dV}{d\phi}-2V +\frac{d\omega}{d\phi} 
\left( \dot{\phi} \right)^2 \right] \nonumber \\
&& \nonumber \\
&& +\frac{K}{a^2} -\frac{8\pi }{\left( 2\omega+3 \right)\phi }\left[
\left( \omega+2\right) \rho^{(m)} +\omega P^{(m)} \right] \;,
\label{6}
\end{eqnarray}

\be \label{7}
\ddot{\phi} +\left( 3H +\frac{1}{2\omega+3}\, 
\frac{d\omega}{d\phi} \right) \dot{\phi} =
\frac{1}{2\omega+3} \left[
2V-\phi \, \frac{dV}{d\phi} +8\pi 
\left( \rho^{(m)} -3P^{(m)} \right) \right] \;,
\ee
where an overdot denotes differentiation with respect to the comoving time $t$, 
$H\equiv \dot{a}/a$ is the Hubble parameter, and $\rho^{(m)}$ and $P^{(m)}$ 
are the energy density and pressure of the material fluid, respectively.
The latter satisfies the conservation equation
\be \label{8}
\dot{\rho}^{(m)} +3H \left( \rho^{(m)}+ P^{(m)} \right) =0 \;.
\ee
We assume that the matter fluid has equation of state
\be 
P^{(m)}=\left( \gamma -1 \right) \rho^{(m)}
\ee
with $\gamma $ a constant. Then eq.~(\ref{8}) is immediately integrated to yield
\be 
\rho^{(m)}=\frac{ \rho_0}{a^{3\gamma}} 
\ee
(with $\rho_0$ a constant), a fact that is used in the following sections. 
In the general case, the dynamical variables are $a$ and $\phi$, 
the dynamical equations 
(\ref{5}) and (\ref{7}) are of second order in $a$ 
and $\phi$ (there are only two independent 
equations in the set 
(\ref{5})-(\ref{7})), and the Hamiltonian constraint reduces the dimensionality 
of the phase space $\left( a, \dot{a}, \phi, \dot{\phi} \right)$ to three,
 as noted in Ref.~\cite{SG} for the special case of Brans-Dicke theory. In 
many situations
it is possible to perform a further 
dimensional reduction of the phase space. First one 
notes that in vacuum and for $K=0$ the scale factor $a$ only appears in
eqs.~(\ref{5})-(\ref{7}) through the combination $H\equiv \dot{a}/a$. One can 
then use the variables $H$ and $\phi$ and the Hamiltonian constraint to reduce 
the dimensionality of the phase space $\left( H, \phi, \dot{\phi} \right) $ to
two. It is appropriate to use the Hubble parameter because it is one of the 
cosmological observables.

It is sometimes useful to rewrite eqs.~(\ref{5})-(\ref{7}) using the conformal
time $\eta$ defined by $dt=a \, d\eta$ and the new variable
\be \label{9}
x\equiv \frac{a'}{a} =aH \;,
\ee
where a prime denotes differentiation with respect to $\eta$. In terms of $x$ 
and $\eta$ one has
\begin{eqnarray} 
x^2 & = & -x\left( \frac{\phi'}{\phi} \right) +\frac{ \omega(\phi)}{6} \left( 
\frac{\phi'}{\phi} \right)^2 +\frac{a^2 V(\phi)}{6\phi} -K 
+\frac{8\pi \rho_0}{3a^{3\gamma-2} \phi} \;,
\label{10}  \\
&& \nonumber \\
 x' & = & x^2 -\frac{\omega}{2} \left( \frac{\phi'}{\phi} 
\right)^2 
+2x \left( \frac{\phi'}{\phi} \right)+\frac{1}{2\left(2\omega+3 \right) \phi}
\left[ a^2 \left( \phi \,\frac{dV}{d\phi} -2V \right) +
\left( \phi' \right)^2 \, \frac{d\omega}{d\phi} \right] \nonumber \\
&& \nonumber \\
&& +K
-\, \frac{8\pi \rho_0 \left( \omega\gamma +2\right)}{\left( 2\omega+3 \right)
 \phi a^{3\gamma-2} } \;.
\label{11}
\end{eqnarray}
In the rest of  this paper we consider various specifications of the 
general scalar-tensor theory (\ref{2}) corresponding to various 
combinations of the scalar field potential and of the cosmic fluid.


\section{Vacuum, free Brans-Dicke-like field, and any $K$  }
\setcounter{equation}{0}

In the case in which no material fluid is present and $V\equiv 0$, 
eq.~(\ref{10}) can be written as
\be \label{19}
\omega \left( \phi'\right)^2 -6x\phi\phi' -6\phi^2 \left(K+x^2\right)=0 \;.
\ee
Provided that $\omega \neq 0$, this is a second degree algebraic equation for 
$\phi'$ with reduced discriminant
\be \label{20}
\frac{\Delta_1}{4}= 3\phi^2 \left[ \left( 2\omega +3 \right)x^2 +2\omega K \right]
\ee
and with solutions
\be \label{21}
\phi'_{\pm}\left( x, \phi \right)= \frac{3\phi}{\omega} 
\left( x\pm \sqrt{ {\cal F}_1 \left( x, \phi \right)} \, \right) \;,
\ee
\be \label{22}
{\cal F}_1 \left( x, \phi \right) =
\frac{1}{3} \left\{  \left[ 2\omega(\phi) +3\right]x^2 
+2K \omega(\phi) \right\} \;.
\ee
If $ \omega>0$ and $K=0$ or $+1$ it is always ${\cal F}_1 \geq 0$. If 
$K=-1$ or $\omega <0$, there can be regions corresponding to ${\cal F}_1<0$ which 
are forbidden to the dynamics. 
The use of eq.~(\ref{11}) yields
\be \label{23}
x'_{\pm} \left( x, \phi \right) = x^2 + \frac{9}{2\omega} 
\left( x \pm \sqrt{ {\cal F}_1} \, \right)^2
\left[ \frac{\phi}{\omega \left( 2\omega+3 \right)}
\, \frac{d\omega}{d\phi} -1 \right] 
+ \frac{6x}{\omega} \left( x\pm \sqrt{ {\cal F}_1}
 \right) +K \;.
\ee
$x'$ and $\phi'$ are determined once the values of $x$ and $\phi$ are 
given. In general there are two values of both $x'$ and $\phi'$ for a given pair 
$\left( x, \phi \right) $, corresponding to the 
upper and lower sign in eqs.~(\ref{23}) and (\ref{21}), respectively. 
This property
 corresponds to the geometry of the phase space which is a two-dimensional
surface embedded in the $\left( x,x', \phi, \phi' \right)$ space
and consists of two sheets, each of which corresponds, respectively, to
 the upper or lower sign in eqs.~(\ref{21}) and
(\ref{23}) -- we can call them ``upper sheet'' and ``lower sheet''. 
The $\left( x, \phi \right) $ 
plane is not the phase space but is a projection of it. 
As a consequence, the study of the $\left( x, \phi \right)$ plane can show 
projections of the orbits intersecting each other. Of course the true 
orbits living in the curved energy surface do not cross, due to the 
uniqueness of the solutions of the Cauchy problem for the system 
(\ref{5})--(\ref{7}), but projections onto the $ \left( x, \phi \right) 
$ plane of two different orbits located in 
the two different sheets can intersect.

The orbits of the solutions can 
only change sheet (and they don't necessarily change in all the possible 
scalar-tensor scenarios) at points where the two sheets touch each other, 
i.e., where ${\cal F}_1=0$. This set of points forms the 
boundary of the forbidden region
\be 
{\cal B}=\left\{ \left( x, \phi, \phi' \right)\: : \;\; 
{\cal F}_1 \left( x, \phi \right) =0 \right\} \;.
\ee
On points of ${\cal B}$ one has 
$x'_{+}=x'_{-}  $ and 
$\phi'_{+}=\phi'_{-} =3\phi \, x/ \omega $. 
It should be noted that 
the phase space reduces to a plane instead of a
 curved two-surface when ${\cal F}_1$ vanishes
identically, i.e., if $K=0$ and $\omega=-3/2$.

In the case of vacuum, $ V\equiv 0$, and $K=0$ the dimensional reduction 
proceeds 
by using comoving time and the variables $H$ and $\phi$ -- see Sec.~4 for 
an example. However 
if $K=\pm 1$ and comoving time
is used, the scale factor cannot be eliminated from the field equations
and the dimensional reduction cannot be achieved. To this purpose one
must use conformal time when $K=\pm 1$. One notes that the potential 
can be neglected
($V\simeq 0$) when the kinetic terms dominate the dynamics.
In this case the evolution takes the universe away from the possibility
of chaos and toward a two-dimensional phase space. If $V( \phi)$ 
dominates over the kinetic terms at late times, as 
in dark energy models, there is in
principle the possibility of chaotic evolution, 
although this is not by all means guaranteed.

The fixed points of the system (\ref{21}) and (\ref{23}) correspond to 
$\left( x', \phi' \right)=\left( 0,0 \right)$; 
eqs.~(\ref{21})--(\ref{23}) then yield either 
$ \phi=0$, which corresponds to infinite gravitational
coupling and is unphysical, or $x=0$ and $\omega \, K=0$. Hence, 

\begin{itemize}

\item For $K=0$ the only fixed point 
is the Minkowski 
space corresponding to $x=aH=0$, which lies on the boundary ${\cal B}$ of 
the region forbidden to the dynamics. 

\item For  $K= \pm 1$ the only fixed points are of the form  $ \left( 
x , \phi \right) = \left( 0, \phi_0 \right) $ where $\phi_0$ is a root of 
the equation $\omega \left( \phi \right)=0$. Such a fixed point  
corresponds again to a Minkowski space lying on  ${\cal B}$. 
If $\omega \neq 0$ everywhere there are no fixed points and, as a  
consequence, there are no limit cycles (a cycle must contain at least a 
fixed point).

\end{itemize}


\section{Vacuum, $V=m^2\phi^2/2$, and $K=0$}
\setcounter{equation}{0}

In the situation of vacuum, a massive scalar $\phi$, and $K=0$ one can 
use again the physical variables $H$ and $\phi$. Eq.~(\ref{5}) is 
rearranged as the 
algebraic equation for $\dot{\phi}$
\be 
\frac{\omega}{6}\left( \dot{\phi} \right)^2-H\phi\dot{\phi} +
\left( \frac{m^2}{12} \, \phi -H^2 \right) \phi^2 =0 
\ee
with discriminant
\be
\Delta_2 \left( H, \phi \right) 
=\frac{\phi^2}{3} \left[ \left( 2\omega+3 \right) H^2 
-\frac{\omega \, m^2}{6} \, \phi \right]
\ee
and solutions
\be
\dot{\phi}_{\pm} \left( H, \phi \right) =\frac{3}{\omega} 
\left( H\phi \pm \sqrt{\Delta_2} \, \right) \;.
\ee
The phase space is again a curved two-dimensional surface 
embedded in the space $ \left( H, \phi, \dot{\phi} \right)$ 
and composed of two sheets. The two sheets join on the boundary of the forbidden
region described by $\Delta_2=0$, which is equivalent to either $\phi=0$ (unphysical) or
\be
H(\phi)= \pm m \, \sqrt{ \frac{ \phi \, 
\omega( \phi)}{ 6 \left[ 2\omega(\phi) +3 \right] } } \;.
\ee
The fixed points $\left( H_0, \phi_0 \right)$ are subject to the 
only restriction $H_0=\pm m\sqrt{ \phi_0 / 12} $ and are de Sitter 
spaces with constant scalar field $\phi_0 \geq 0$ located away from the 
boundary $\Delta_2=0$.
Again, if $K\neq 0$ the scale factor cannot be eliminated and one cannot choose $H$ as dynamical variable. 
For any value of $K$ the wave equation for $\phi$ can be 
integrated yielding either the trivial solution $\phi =$constant (corresponding
 to general relativity with a cosmological constant) or
\be
\int d\phi \, \sqrt{ 2\omega (\phi) +3} =\mbox{const.} \int \frac{dt}{a^3} \;.
\ee
A complete description of the phase space is given by Santos and Gregory 
\cite{SG} 
for Brans-Dicke theory with $\omega=$constant. They use the dynamical 
variables
\begin{eqnarray}
X & \equiv & \sqrt{ \frac{ 2\omega+3}{12} } \, \frac{\phi'}{\phi} \;, \\
&& \nonumber \\
Y & \equiv & \frac{a'}{a} +\frac{\phi'}{ 2 \phi} \;,
\end{eqnarray}
and conformal time $\eta$.
The only difference between the Brans-Dicke case and more general 
scalar-tensor 
theories are the terms in $\dot{\omega}$ in the field equations of 
general scalar-tensor gravity. However, the
dimensional reduction of the phase space cannot be achieved
 by using the variables $X, Y$, and $\eta$.

\section{The case of vacuum, $K=0$, and $V\neq 0$}
\setcounter{equation}{0}

Let us consider again vacuum and a 
general self-interaction potential $V(\phi)$ 
for the Brans-Dicke-like field $\phi$, but let us impose the restriction 
that the universe 
has flat spatial sections ($K=0$). This is the situation
indicated by the Boomerang \cite{Boomerang} and MAXIMA \cite{MAXIMA} 
experiments for our universe. In this section we use 
the physically transparent variables $H$ and $\phi$ and  
comoving instead of conformal time. The dynamical equations reduce to
\be \label{24}
H^2 = -H \left( \frac{ \dot{\phi}}{\phi} \right) 
+\frac{\omega(\phi)}{6} \left( \frac{\dot{\phi}}{\phi} \right)^2
+ \frac{V(\phi)}{6\phi} \;,
\ee

\be \label{25}
\dot{H}= -\frac{\omega(\phi)}{2} \left( \frac{ \dot{\phi}}{\phi} \right)^2 
+2H \left( \frac{\dot{\phi}}{\phi} \right) +\frac{1}{2\left( 2\omega+3 \right) \phi} 
\left[ \phi \, \frac{dV}{d\phi} -2V 
+\frac{d\omega}{d\phi} \left( \dot{\phi} \right)^2 \right]\;,
\ee

\be \label{26}
\ddot{\phi} +\left( 3H 
+\frac{1}{2\omega+3 } \, \frac{d\omega}{d\phi} \right) \dot{\phi} =
\frac{1}{2\omega+3} \left( 2V - \phi \, \frac{dV}{d\phi} \right) \;.
\ee
When $\omega \neq 0$ eq.~(\ref{24}) can be rearranged as the algebraic 
equation for $\dot{\phi}$
\be \label{27}
\omega \left( \dot{\phi} \right)^2 -6H\phi\dot{\phi} +\left( V-6H^2 \phi \right) 
\phi =0 
\ee
with reduced discriminant
\be \label{28}
{\cal F}_2 \left( H, \phi \right) =\left[ 3 \left( 2\omega+3 \right) H^2 
\phi 
-\omega V \right] \phi
\ee 
and solutions 
\be \label{29}
\dot{ \phi}_{\pm} \left( H, \phi \right) = 
\frac{1}{\omega(\phi)} \left[ 3H \phi \pm 
 {{\cal F}_2}^{1/2} \left( H, \phi \right)   \right] \;.
\ee
The phase space is again a two-dimensional surface composed of two 
sheets corresponding to the 
lower and upper signs in eq.~(\ref{29}). Figures 1--9 illustrate the 
two--sheeted structure of the phase space for particular choices of 
$\omega ( \phi )$ and $V( \phi) $. In general there is no guarantee 
that ${\cal F}_2 \geq 0$ and there will be regions of the space 
$\left( H, \phi, \dot{\phi} \right) $ corresponding to ${\cal F}_2 <0$ 
forbidden to the
orbits of the solutions. 
If $K\neq 0$ the reduction of the phase space to two dimensions cannot be performed
because the scale factor appears in the expression $K/a^2$ in eqs.~(\ref{5})  
and (\ref{6}) and not in the form $H=\dot{a}/a$.

The phase space is flat if ${\cal F}_2 =0$, i.e., if $\omega=-3/2$ and $ 
V \equiv 0 $. In this case one has 
\be 
\dot{\phi}+2H\phi=0
\ee
and either $\phi$ is identically zero (a physically unacceptable solution)
or $\phi \propto a^{-2}$. Then the Hamiltonian constraint is automatically 
satisfied and $\dot{H}=-H^2$, which yields
the coasting universe with scale factor $a=a_0 \left( t-t_0 \right)$.

In the general situation with $V\neq 0$ studied in this section, the 
equilibrium points correspond to 
$ \left( H, \phi \right)=\left( H_0, \phi_0 \right) $ with $H_0$ and $\phi_0$ 
constants and must satisfy 
\be
H_0=\pm \sqrt{ \frac{V_0}{6\phi_0}} \;, \;\;\;\;\;\;\;\;\;\;\;\;\;\;
\phi_0 V_0'-2V_0=0 \;,
\ee
where $V_0 \equiv V( \phi_0) $ and $\left.  V_0'=dV/d\phi \right|_{\phi_0}$.


\section{The case $\gamma=2/3$ and $ V \equiv 0$}
\setcounter{equation}{0}

Another special case corresponds to the choice $\gamma=2/3$ (or $P=-\rho/3$) 
and $V\equiv 0$. In terms of the variables $x$ and $\phi$ and of conformal time the field equations reduce to
\be \label{12}
x^2=-x\left( \frac{ \phi'}{\phi} \right) +\frac{ \omega(\phi)}{6} \left( 
\frac{\phi'}{\phi} \right)^2 -K +\frac{8\pi \rho_0}{3\phi} \;,
\ee
\be \label{13}
x'=x^2-\frac{\omega(\phi)}{2} \left( \frac{\phi'}{\phi} \right)^2 
+2x\frac{\phi'}{\phi} +K 
+\frac{\left( \phi'\right)^2}{2\left( 2\omega+3 \right)\phi}\, \frac{d\omega}{d\phi} 
-\frac{A}{\phi} \;,
\ee  
where
\be 
A=\frac{ 16\pi \rho_0 \left( \omega+3 \right)}{3\left( 2\omega +3 \right)} \;.
\ee
Eq.~(\ref{12}) can be rewritten as the algebraic equation for $\phi'$
\be
\frac{\omega}{6} \left( \phi'\right)^2 -x\phi\phi'+
\left( \frac{8\pi \rho_0 \phi}{3} -K\phi^2-x^2\phi^2 \right)=0
\ee
with discriminant
\be
\Delta_3 \left( x, \phi \right)=\frac{\phi}{3} \left[ \left( 2\omega+3\right) x^2\phi 
-2\omega \left( \frac{8\pi \rho_0}{3} -K\phi \right)\right]
\ee
and solutions
\be
\phi'_{\pm}\left( x, \phi \right)=\frac{3}{\omega} \left( 
x\phi \pm \sqrt{\Delta_3 \left( x, \phi \right)} \right) \;,
\ee
while eq.~(\ref{13}) yields 
\begin{eqnarray}
x' & = & x^2-\frac{9}{2\omega\phi^2} \left( x\phi \pm \sqrt{ \Delta_3} \,
\right)^2 
+\frac{6x}{\omega\phi} \left( x\phi \pm \sqrt{\Delta_3} \, \right) +K
\nonumber \\
&& \nonumber \\
&+ & \frac{9}{2\omega^2 \left(2\omega+3 \right)\phi} \,
\frac{d\omega}{d\phi}\left( x\phi \pm \sqrt{\Delta_3} \, 
 \right)^2 -\frac{A}{\phi} \;.
\end{eqnarray}
The geometry of the phase space is again that of two sheets attached along the 
boundary $\Delta_3 \left(  x, \phi \right) =0$ 
of a forbidden region and embedded in the space $\left( x, \phi, \phi' \right)$.
There are no fixed points in this case. In fact, the fixed points would 
correspond to constant $x_0$ and $\phi_0$ and it is easy to see that 
eqs.~(\ref{12}) and (\ref{13}) are incompatible when 
$x$ and $\phi$ are both constant.

\section{Discussion and conclusions}
\setcounter{equation}{0}

The geometry and dimensionality of the phase space of FLRW scalar-tensor cosmology,
in which gravity is described by the action (\ref{2}), can be quite complicated. 
In general one has two independent equations for scalar-tensor cosmology, the 
wave equation (\ref{7}) for the scalar $\phi$ and eq.~(\ref{6}) for the 
scale factor $a$. Since these are second order equations one is naturally lead to 
consider the phase space $\left( a, \dot{a}, \phi, \dot{\phi} \right)$. 
The Hamiltonian constraint (\ref{5}) 
provides a first integral and the orbits of the solutions
lie on a hypersurface in this space. Hence in general the phase space is a 
three-dimensional curved energy hypersurface embedded in the four-dimensional
space $\left( a, \dot{a}, \phi, \dot{\phi} \right) $. When the spatial sections are 
flat ($K=0$) the scale factor only enters the field equations through 
the Hubble parameter and
it is convenient to choose as physical variables $H$ and $\phi$ instead of 
$a$ and $\phi$.  The Hamiltonian constraint (\ref{5}) then forces the 
traiectories of the 
solutions to unfold in a two-dimensional hypersurface  embedded in the 
three-dimensional space $\left( H, \phi, \dot{\phi} \right)$. 
However, there are also situations in which the universe has curved spatial 
sections ($K=\pm 1$) and one can again reduce the phase space to a 
two-dimensional
surface embedded in a three-dimensional space. Table~1 summarizes the 
situations 
known in the literature and the results of the present paper. The latter are 
derived without specifying the coupling function $\omega(\phi)$ 
and, in some cases, also for arbitrary scalar field potential $V(\phi)$.
The stationary points of the dynamical system are 
determined in this general situation.

There is a simplification in the equations of scalar-tensor cosmology when
the potential reduces to a mass term -- previous works \cite{KE,SG} have taken 
advantage of this property. When a general scalar-tensor theory is considered
instead of Brans-Dicke theory the time-dependence of $\omega(\phi)$ makes 
the dynamics more interesting even 
for vacuum, $K=0$ and  a free scalar field. 
Even when the phase space is two-dimensional its geometry is usually 
non-trivial.
In general the phase space accessible to the trajectories of the 
solutions is 
composed of two sheets attached to each other along the boundary 
${\cal B} $ of a  region forbidden to the dynamics. The orbits of the solutions
can change sheet only by passing through points of ${\cal B}$. There are 
however scenarios in which the orbits are forced to stay in one sheet or 
the other but cannot change sheet -- the sheet in which they live is 
decided by the initial conditions. The plane
$\left( H, \phi \right)$ (or $\left( x, \phi \right)$) is only a projection of 
the true 
curved phase space  and projections of the orbits can intersect in this plane 
(apart from the degenerate cases in which this plane reduces to the 
true phase space). In all the situations considered in this paper the 
fixed points of the dynamical system (when they exist) are found to be de 
Sitter spaces possibly degenerating into Minkowski spaces. Their stability 
is studied in Ref.~\cite{deSitter}.

A detailed discussion of the dynamics requires the specification of the functions 
$\omega(\phi)$ and $V(\phi)$. Although many forms of $V(\phi)$ have been 
considered in the literature on high energy physics and cosmology,
there are very few suggestions about the form of the coupling function 
$\omega( \phi)$, and they are usually dictated by purely 
mathematical considerations
 such as 
the ease of finding  
exact cosmological solutions (cf. Ref.~\cite{mybook}). The clarifications 
on the dimensionality and geometry of the phase space made here will be useful
in detailed studies fixing the form of $\omega(\phi) $ and $V(\phi)$.

\section*{Acknowledgments}

It is a pleasure to thank Prof. Werner Israel for the hospitality at 
the University of Victoria, where this paper was completed. This 
work was supported by the Natural Sciences and Engineering Research 
Council of Canada ({\em NSERC}). 


\clearpage

\begin{center} {\bf Table caption} \end{center}

\noindent {\bf Table~1:} Situations in which the geometry and dimension of 
the phase space 
are known (BD~$\equiv$~Brans-Dicke theory, NMC~$\equiv $~nonminimally 
coupled scalar field theory).

\vskip2truecm

\begin{center} {\bf Figure captions} \end{center}

\noindent {\bf Figure 1:} The upper sheet of the phase space of 
Brans-Dicke theory described by 
$\omega= 10 $ and scalar field potential 
$V(\phi)=m^2 \phi^2/2$ (for unit 
mass $m$).  \\\\
\noindent {\bf Figure 2:} The lower sheet corresponding to the situation 
of fig.~1.\\\\
\noindent {\bf Figure 3:} The phase space of Brans-Dicke theory with 
massive scalar 
resulting from the two sheets of 
fig.~1 and fig.~2 glued together. The region forbidden to the dynamics 
appears as a hole that cannot be penetrated by the orbits of the 
solutions.\\\\
\noindent {\bf Figure 4:} The upper sheet of the phase space of the 
scalar-tensor theory described by 
$\omega= \left| \phi -1 \right|^{-1} $ and scalar field potential 
$V(\phi)= m^2  \phi^2 /2  $ (for unit mass $m$).  \\\\  
\noindent {\bf Figure 5:} The lower sheet corresponding to the situation 
of fig.~4.\\\\
\noindent {\bf Figure 6:} The phase space resulting from the two sheets of 
fig.~4 and fig.~5 glued together. There is a forbidden region and the two 
sheets touch each other only at the boundary of this region.\\\\
\noindent {\bf Figure 7:} The upper sheet of the phase space of the 
scalar-tensor theory described by 
$\omega= \left( \phi -1 \right)^{-4} $ and scalar field potential 
$V(\phi)=V_0 \,\cos \left(  \phi/m \right) $ (for $m$ and $V_0$ set to 
unity).  \\\\  
\noindent {\bf Figure 8:} The lower sheet corresponding to the situation 
of fig.~4.\\\\
\noindent {\bf Figure 9:} The phase space resulting from the two sheets of 
fig.~4 and fig.~5 glued together. There is a forbidden region and the two 
sheets touch each other only at the boundary of this region.

\clearpage

\thispagestyle{empty}
\noindent
\begin{table}
{\small {\small


\begin{tabular}{|c|c|c|c|c|c|}

\hline 
$ 
\begin{array}{c} \\
\mbox{Scalar-tensor}\\
\mbox{theory}\\ \\ \end{array}$ 
& 
cosmic fluid   
& 
 $ V\left( \phi \right) $ 
& 
$\begin{array}{c} 
\mbox{curvature}\\
 \mbox{index} \end{array} $ 
&
reference
&
phase space\\
\hline 
\hline 

$\begin{array}{c} \\  \mbox{BD} \\ \\ \end{array} $  
& absent & $ V= \frac{m^2\phi^2}{2}$ & any $K$  & \cite{SG} & 2-D flat \\

\hline $\begin{array}{c} \\ \mbox{BD} \\ \\ \end{array} $ 
& $P=-\rho/3$ & $ V=\frac{m^2\phi^2}{2} $ & any $K$  & this paper & 2-D curved\\

\hline $\begin{array}{c} \\ \mbox{BD} \\ \\ \end{array} $  
& $ \begin{array}{c} \mbox{any} \\ \mbox{perfect fluid}
\end{array} $ 
& 
$ V\equiv 0$ & any $K$  & \cite{KE,HW} & 2-D flat\\

\hline $\begin{array}{c} \\ \mbox{BD} \\ \\ \end{array} $   
& $ \begin{array}{c} \mbox{any} \\
\mbox{perfect fluid} \end{array}$  & $ V=\Lambda \phi$ & $K=0$  & \cite{KE}
 & 2-D flat \\

\hline $\begin{array}{c} \\ \mbox{BD} \\ \\ \end{array} $  
& $\begin{array}{c} \mbox{any}\\ \mbox{perfect fluid}\end{array} $ 
 & any $V(\phi)$ & any $K$  & \cite{KE,SG} & 3-D curved\\

\hline $ \begin{array}{c} \\ \mbox{NMC} \\ \\ \end{array} $
  & absent & any $V(\phi)$ & $K=0$  & \cite{Foster}--\cite{SaaetalIJTP}
 & 2-D curved\\

\hline  $\begin{array}{c} \\ \mbox{NMC} \\ \\ \end{array} $  
& absent & any $V(\phi)$ & $K=\pm 1$  & \cite{Foster}--\cite{SaaetalIJTP}
& 3-D curved\\

\hline  $ \begin{array}{c} \\ \mbox{Action} \\ \mbox{(\ref{2})} \\ \\ \end{array} $ 
  & absent & $ V\equiv 0$ & any $K$  & this paper & 2-D 
curved\\

\hline $ \begin{array}{c} \\ \mbox{Action} \\ \mbox{(\ref{2})} \\ \\ \end{array} $ 
  & absent & any $ V(\phi)$ & $K=0$  & this paper & 2-D 
curved\\

\hline $ \begin{array}{c} \\  \mbox{Action} \\ \mbox{(\ref{2})} \\ \\ \end{array} $  
 & absent & any $ V(\phi)$ & $K=\pm 1$  & this paper & 3-D curved\\

\hline
\end{tabular}
}}
\end{table} 

\thispagestyle{empty}

\end{document}